\documentclass[a4paper]{jpconf}
\usepackage{graphicx}
\usepackage{lineno}

\input babarsym.tex
\def\Btag{\ensuremath{B_{\rm tag}}\xspace}  

\def \Bsig{\ensuremath{B_{\rm sig}}\xspace}

\begin{document}
\title{Search for rare \B meson decays at the \babar{} experiment}

\author{R. Cheaib}

\address{On behalf of the \babar{} collaboration\\Department of Physics, University of Mississippi, 108 Lewis Hall, University MS 38677}

\ead{rachac@phys.olemiss.edu}

\begin{abstract}
\b\to\s transitions are flavour-changing neutral current (FCNC) processes that play an important role in the search for physics beyond the Standard Model (SM). Contributions from virtual particles in the loop are predicted to deviate observables, such as the branching fraction, from their SM expectations. Using data from the BaBar experiment, we present the first search for the rare decay \Bp\to\Kp\tautau. The \babar{} results  on the measurement of the angular asymmetries of \B\to\Kstar\ellell, where $\ell=e$ or $\mu$,  are also reported. In addition, using a time-dependent analysis of $\B\to\KS\pip\pim\gamma$, the mixing induced CP-asymmetry for the radiative FCNC decay, $\B\to\KS\rho\gamma$, is measured, along with an amplitude analysis of the $m_{\kaon\pi}$ and $m_{\kaon\pi\pi}$ spectrum.
\end{abstract}

\section{Introduction}

\b\to\s transitions are highly suppressed in the SM and only occur via loop or box diagrams. Using an effective low-energy theory, the Lagrangian for \b\to\s transitions can be separated into two distinct parts: the long distance (low-energy) contributions contained in the operator matrix elements and the short-distance (high-energy) physics described by the Wilson coefficients. Measurements of such rare FCNC \B meson decays are interesting since they can provide experimental constraints on the associated Wilson coefficients and are thus a stringent test of the SM. Furthermore, contributions from new-physics particles, like a charged Higgs \cite{aliev}, can deviate these Wilson coefficients or require the introduction of new opertaor matrix elements, and thus this is an interesting hunt for physics beyond the SM.

The \babar{} experiment~\cite{babar2} collected 424 fb$^{-1}$ of data~\cite{lum} by colliding electrons and positrons at the center-of-mass (CM) energy of the $\FourS$ resonance. The \FourS decays into \BB pair, resulting in more than 479 million \BB events to study and analyze. Using the full \babar{} dataset, a measurement of the \Bp\to\Kp\tautau branching fraction~\cite{ktautau}, \B\to\Kstar\ellell angular asymmetries~\cite{kll} and $\B\to\KS\rho\gamma$ mixing-induced CP-asymmetry~\cite{ksRhoGamma} is performed.

\section{Branching fraction measurement of \Bp\to\Kp\tautau}

\Bp\to\Kp\tautau~\cite{cc} is a FCNC process with a braching fraction in the range $1$-$2 \times 10^{-7}$~\cite{lattice}. It is the third-lepton generation equivalent of \B\to\kaon\ellell, where $\ell= e$ or $\mu$. The large mass of the $\tau$ lepton may provide improved sensitivity to new-physics contributions as compared to its light lepton counterparts ~\cite{ktautauTheory1,ktautauTheory2}. For instance, in two-Higgs-doublet-models~\cite{aliev}, the Higgs-lepton-lepton vertex is proportional to $m^{2}_{\tau}$ and thus contributions to the total decay rate, as well as other observables, can be significant. 
The branching fraction of \Bp\to\Kp\tautau is measured by exclusively reconstructing one \B meson, referred to as the \Btag, in the \FourS\to\BB decay using hadronic modes, and then examining the rest of the event for the \Bp\to\Kp\tautau signal. This technique is referred to as the hadronic \Btag reconstruction, and is ideal for decays with missing energy. With exclusive reconstruction of the \Btag, the four-vector of the other \B, the \Bsig, can also be fully determined and thus the kinematics of the event are better constrained. The $\tau$ is required to decay only via leptonic modes: $\taum\to\en\nueb\nut$ or $\taum\to\mun\numb\nut$, and thus there are three possible final states: \epem, \mumu or \ep\mun with their associated neutrinos.

\begin{figure}
\begin{center}
\includegraphics[height=4cm,width=8cm]{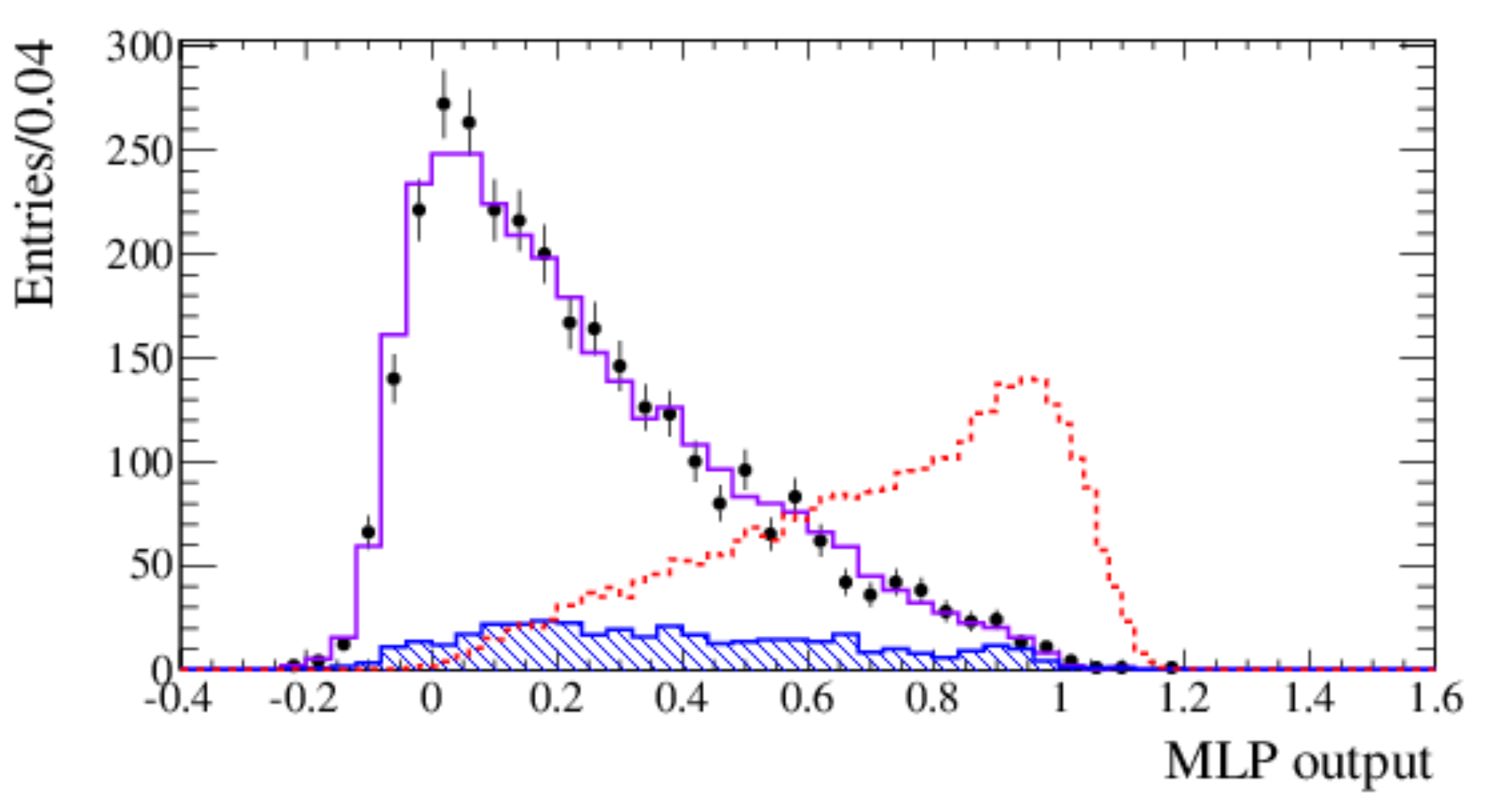}\\
\caption{ MLP output distribution for the \Bp\to$K^{+}$\tautau analysis. The signal MC distribution is shown (dashed) with arbitrary normalization, along with the data (points) and the expected combinatorial (shaded) plus \mes-peaking (solid) background contributions.}
\label{fig2}
\end{center}
\end{figure}

To select for \Bp\to\Kp\tautau, every event is required to have exactly one properly-reconstructed charged \Btag with an energy substituted mass, \mes, that lies within the range of the mass of a \B meson. The \mes of a \Btag candidate is given by $\mes=\sqrt{E_{\rm CM}^{\ 2}- {\vec{p}_{\Btag}}^{ \ 2}}$, where $E_{\rm CM}$ is half the total colliding energy and $\vec{p}_{\Btag}$ is the 3-momentum of the reconstructed \Btag, in the CM frame. 
Furthermore, \Bp\to\Kp\tautau events are required to have nonzero missing energy, which is calculated by subtracting all signal side tracks and clusters from the \Bsig four-vector. A signal event is also required to have 3 tracks, one satisfying the particle identification (PID) criteria of a \kaon and two of an electron or muon. In addition, the presence of massive $\tau$ leptons imposes an upper limit on the \kaon momentum, which restricts the $s_{B}$ distribution, given by $s_{B}=(p_{\Bsig}-p_{\kaon})^{2}/m^{2}_{\B}$,  of signal events to higher values as compared with background events. Therefore, a requirement of $s_{B}>0.45$ is applied.
At this point, the main source of background is from combinatorial events with semi-leptonic charmed \B decays, such as $\B\to D^{(*)}\ell\nul, D^{(*)}\to\kaon\ell\nul$.  To suppress this background, a multi-layer perceptron (MLP) neural network, consisting of eight discriminating variables such as the angle between the lepton and the oppositely charged kaon in the \tautau rest frame , is used. The neural network is then trained and tested for each of the three signal channels, and the combined MLP output is shown in Fig. \ref{fig2}. The final step in the signal selection is a requirement on the MLP output for each signal channel.

The data yield, after the MLP output cut, is determined seperately for each of the three signal channels and then combined into one \Bp\to\Kp\tautau result. No significant excess is observed in the \epem or \mumu channels, while a 3.7 $\sigma$ excess is obesrved in the \ep\mun channel. Examination of the input and output distributions of the \ep\mun channels does not show any clear evidence of signal-like behaviour or mis-modelling of the background. Furthermore, the combined excess for all three modes is less than 2 $\sigma$. In the absence of signal, the combined upper limit at the 90\% confidence level is measured to be $\BR(\Bp\to\Kp\tautau)<2.6 \times 10^{-3}$. This is the first search for  \Bp\to\Kp\tautau.

\section{Angular asymmetries in \B\to\Kstar\ellell}

\B\to\Kstar\ellell is also a FCNC process, with an amplitude expressed in terms of hadronic form factors and the $C_{7}, C_{9}, $ and $C_{10}$ Wilson Coefficients\cite{kll}. The angular distributions of \B\to\Kstar\ellell, specifically the \Kstar longitudinal polarization, $F_{L}$, and the forward-backward asymmetry, $A_{FB}$,  are notably sensitive to physics beyond the SM~\cite{theoryKll1}-\cite{theoryKll2} and have been previously measured by various experiments~\cite{Bellekll}-\cite{ATLASkll}.  

For any given value of the $q^{2}$,  the kinematic distribution of the \B\to$\Kstar$\ellell decay products can be expressed in terms of three distinct angles: $\theta_{K}$, the angle between the \kaon and the \B in the \Kstar rest frame, $\theta_{l}$, the angle between the lepton and the \B in the \ellell rest frame, and $\phi$, the angle between the \ellell and $\kaon\pi$ decay planes in the \B rest frame~\cite{kll}. After integrating out $\phi$ and $\theta_{l}$, $F_{L}$ can be determined using a fit to $\cos\theta_{K}$ of the form \mbox{$\frac{1}{\Gamma(q^{2})}\frac{d\Gamma}{d(\cos\theta_{K}}= \frac{3}{2}F_{L}(q^{2})\cos^{2}\theta_{K}+\frac{3}{4}(1-F_{L}(q^{2}))(1-cos^{2}\theta_{K})$}. Similarly, $A_{FB}$ can be extracted after integrating over $\phi$ and $\theta_{K}$ using a fit to $\theta_{l}$ of the form \mbox{$\frac{1}{\Gamma(q^{2})}\frac{d\Gamma}{d(\cos\theta_{l}}=\frac{3}{4}F_{L}(q^{2})(1-cos^{2}\theta_{l})+\frac{3}{8}(1-F_{L}(q^{2}))(1+cos^{2}\theta_{l})+A_{FB}(q^{2})\cos\theta_{l}$}~\cite{kllEqs}.

To measure $F_{L}$ and $A_{FB}$, \B\to\Kstar\ellell  signal events are reconstructed in one of the following final states: \Kstarp(\to\KS\pip)\mumu, \Kstarz(\to\Kp\pim)\mumu, \Kstarp(\to\Kp\piz)\epem, \Kstarp(\to\KS\pip)\epem, \Kstarz(\to\Kp\pim)\epem. Each \Kstar candidate is required to have an invariant mass ranging between $0.72$ and $1.10\gevcc$, and is combined with a pair of leptons, each with an individual momenta greater than 0.3 \gevc. The \mes and $\Delta E$ of the resulting \B candidate is then determined and used to separate between signal and background events. Here, $\Delta E=E^{*}_{B}-E_{CM}/2$, where $E^{*}_{B}$ is the CM energy of the \B and $E_{CM}$ is total CM energy.

The main source of background is from semileptonic \B and $D$ decays, as well as continuum background events with random combinations of leptons. Eight bagged decision trees (BDT) are trained to suppress these \BB and \qqbar backgrounds and a final requirement on $\Delta E$ and $L_R$ is applied at the end of the signal selection. Here, $L_R$ is a likelihood ratio which uses the output of the \BB BDT to determine how likely a given event is signal vs background.

\begin{figure}
\begin{center}
\includegraphics[height=5cm]{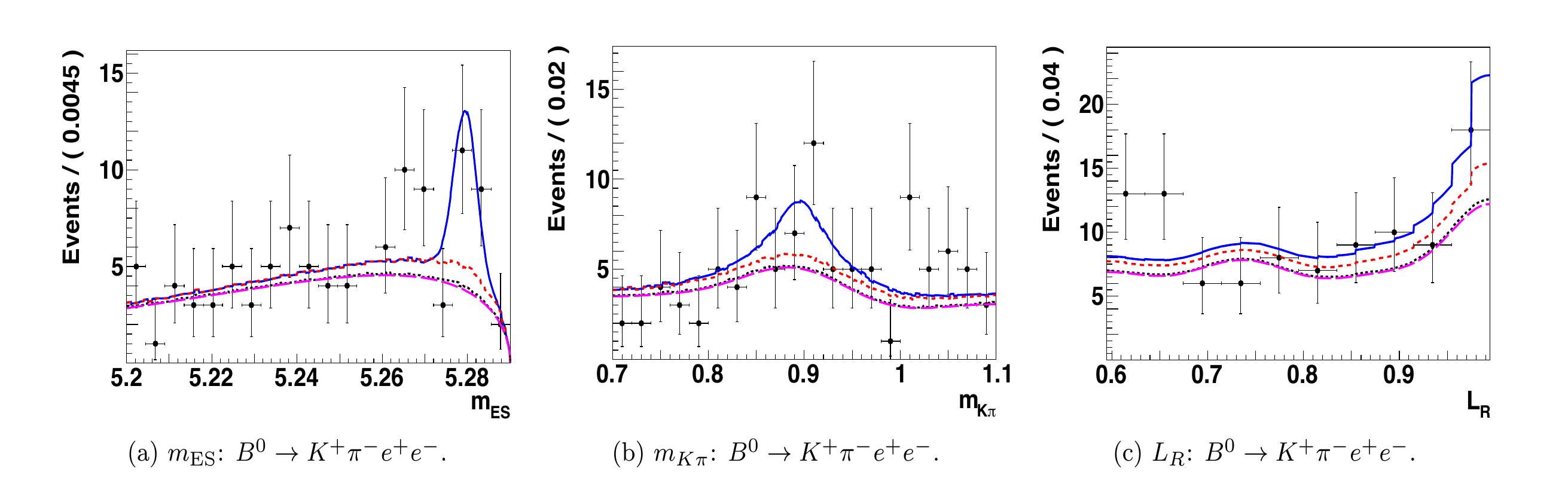}
\caption{ 3-D fit projections for \Bz\to\Kp\pim\epem in $q_5^2$. Each event class is shown: combinatorial (magenta long dash) and charmonium background events (black dots), crossfeed signal events (red short dash) and total pdf (solid blue).}
\label{fig4}
\end{center}
\end{figure}
\begin{figure}
\begin{center}
\includegraphics[height=4cm]{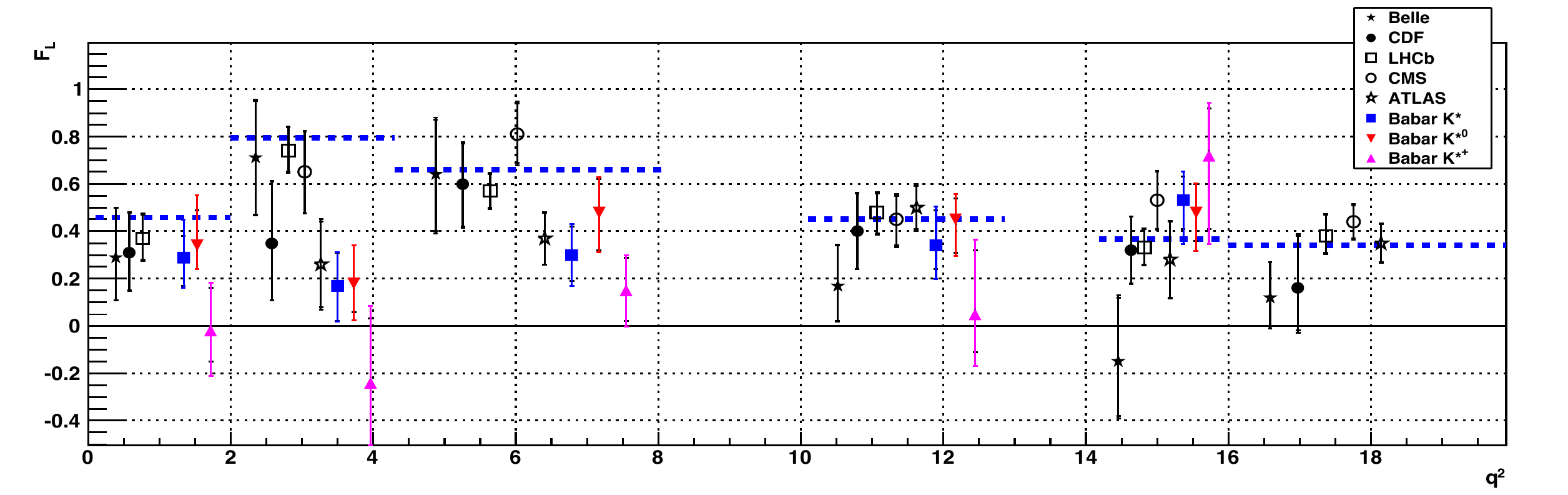}\\
\includegraphics[height=4cm]{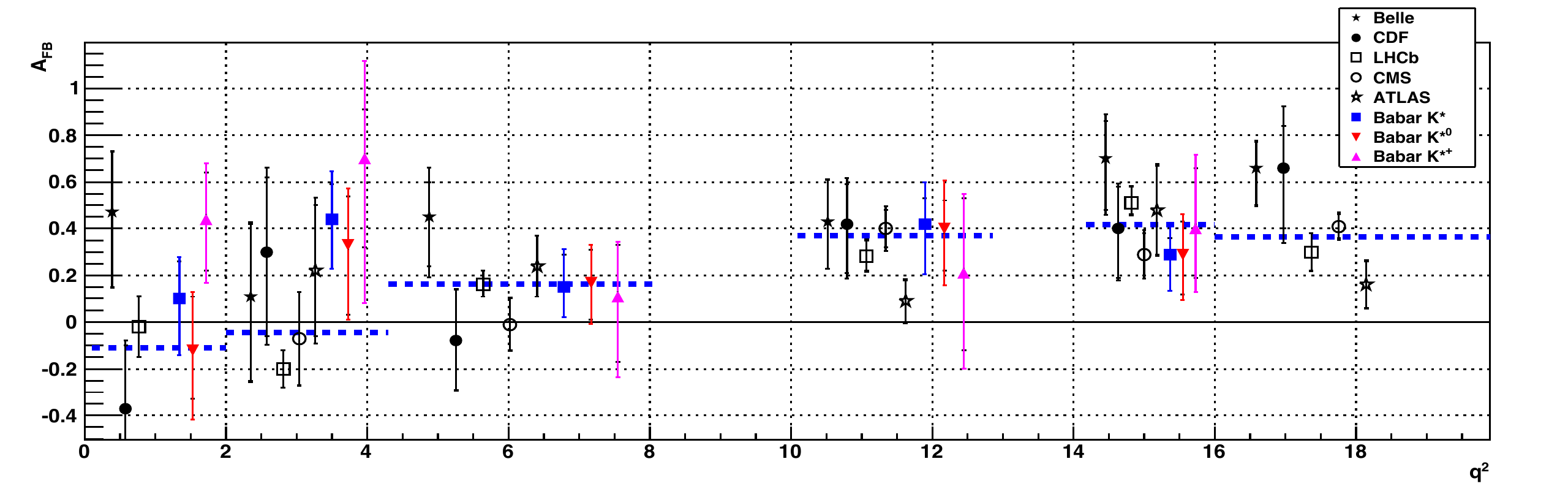}
\caption{ $F_L$ and $A_{FB}$ results for charged (magenta filled pointing-up triangle), neutral (red filled down-pointing triangle) and total \B\to\Kstar\ellell (blue filled square) in disjoint $q^2$ bins. The SM expectations are  shown as blue dashed lines along with results from other experiments: Belle~\cite{Bellekll} (black-filled star), CDF~\cite{CDFkll} (black-filled circle), LHCB~\cite{LHCbkll} (black open square), CMS~\cite{CMSkll} (black open circle), and ATLAS~\cite{ATLASkll} (black open star). }
\label{fig5}
\end{center}
\end{figure}

To extract the angular observables, the $q^{2}$ spectrum is divided into five disjoint bins ($q_1-q_5$) of varying size, and an additional bin $q_0$, ranging between 1.0 and 6.0 GeV$^{2}$/$c^{4}$. An initial unbinned maximum likelihood fit to the \mes, $m(\kaon\pi)$, and $L_R$ spectrums is performed to fix the normalizations and shapes of all probability density functions (pdfs) dependent on these three variables. Then, for each mode and each $q^2$ bin, the 3-D likelihood fit is used to fix the normalizations of events with $\mes>5.27\gevcc$. Third, $\cos\theta_{K}$ is added as a fourth dimension to the likelihood function, and four-dimensional likelihoods are defined for each signal mode and each $q^2$ bin with $F_L$ as the only free parameter. Finally, the fitted value of $F_L$ is then used as input to a similar 4-D fit, where $\cos\theta_l$ has been added as a fourth dimension instead of $\cos\theta_K$. The pdfs in the likelihood fit are defined for five different event classes: true signal events, crossfeed signal events, combinatorial backgrounds, backgrounds from charmonium decays, and finally backgrounds from hadronic decays which are only prominent for \mumu modes. Fig.~\ref{fig4} shows the initial 3-D fit projections for \Bz\to\Kp\pim\epem in $q^{2}_5$ with the different event classes. $F_L$ and $A_{FB}$ are  extracted in each $q^2$ bin for the charged, \Bp\to\Kstarp\ellell, neutral, \Bz\to\Kstarz\ellell, and total \B\to\Kstar\ellell modes. The results are shown in Fig.~\ref{fig5} along with previous ones from Belle ~\cite{Bellekll}, CDF~\cite{CDFkll}, LHCb~\cite{LHCbkll}, and CMS~\cite{CMSkll}.

As can be readily seen, the \Bz\to\Kstarz\ellell results show good agreement with the SM expectations and other experiments. For the charged mode, the value of $F_L$ is relatively small in the low $q^2$ region and thus exhibits tension with the SM expecation. An additional angular observable $P_2$ is defined such that $P_2= (-2/3)\times A_{FB}/(1-F_L)$. $P_2$ has diminished theoretical uncertainty and thus higher sensitvity to non-SM contributions. The tension at low $q^2$ is still found in the $P_2$ distribution and can be a hint of new physics, specifically the results are consistent with the existence of right-handed currents. This is the first measurement of angular asymmetries in \Bp\to\Kstarp\ellell.

\section{Mixing induced CP-asymmetry, $S_{fCP}$, in $\B\to\KS\rho\gamma$}

In the SM, the photon emitted in $\b\to\s\gamma$ transitions is predominantly left-handed, with contamination from right-handed photons suppressed by a factor of $m_s/m_b$~\cite{btos1}. This implies that \Bz (\Bzb) mesons decay predominantly to right-handed (left-handed) photons and the mixing-induced CP-asymmetry in $\B\to f_{CP}\gamma$ decays is expected to be small. However, various new physics models \cite{btosTheory1,btosTheory2} introduce enhanced contributions from right-handed photons and thus alter the prediction of a small CP-asymmetry. 

In this analysis, the mixing-induced CP-asymmetry of $\Bz\to\KS\rho\gamma$, $S_{\KS\rho\gamma}$, is measured using a time-dependent analysis of $\Bz\to\KS\pip\pim\gamma$. Due to the large natural width of the $\rho$ meson, there is an irreducible contribution from the non-CP eigenstate $\Bz\to\Kstarpm(\to\KS\pipm)\pimp\gamma$ which affects $S_{\KS\rho\gamma}$, and thus a dilution factor, $D_{\KS\rho\gamma}\equiv S_{\KS\pip\pim\gamma}/S_{\KS\rho\gamma}$, must be determined. Here, $S_{\KS\pip\pim\gamma}$ is the effective value of the mixing-induced CP asymmetry  for the full $\Bz\to\KS\pip\pim\gamma$ dataset. To determine $D_{\KS\rho\gamma}$, an amplitude analysis of the $m_{\kaon\pi}$ spectra must be performed. Given the small number of events expected in the $\Bz\to\KS\pip\pim\gamma$ sample, the amplitudes of the resonant modes are determined from the charged $\Bp\to\Kp\pip\pim\gamma$ mode instead and then extracted, under the assumption of isospin asymmetry, to the neutral mode.  Furthermore, because the decay to the $\Kp\pip\pim\gamma$ final state proceeds in general through three-body resonances first which then further decay into their $\Kstar\pi$ or $\kaon\rho$ components, it is necessary to determine the three-body resonance content of the $m_{\kaon\pi\pi}$ spectrum as well.

$\Bp\to\Kp\pip\pim\gamma$ events are reconstructed from one high energy photon with \mbox{$1.5<E_{\gamma}<3.5$} \gev, two oppositely-charged pions, and one charged kaon. These are combined to form a \B candidate, whose \mes should lie with 5.20 and 5.92 \gevcc and $|\Delta E|<0.200 \gev$. A Fisher discriminant, consisting of six discriminating variables, is trained to suppress continuum background events. Furthermore, to reduce backgrounds from photons that originate from \piz or $\eta$ decay, a likelihood ratio, $L_R$, is constructed. To extract the $\Bp\to\Kp\pip\pim\gamma$ yield, an unbinned extended maximum likelihood fit to the \mes, $\Delta E$, and Fisher discriminat output $F$ is performed.  The resulting branching fraction is measured to be \mbox{$\BR(\Bp\to\Kp\pip\pim\gamma)=(24.5\pm0.9\pm1.2)\times 10^{-6}$}.

The $m_{\kaon\pi\pi}$ spectrum is then extracted from the maximum likelihood fit, and modeled as the coherent sum of five kaonic Breit-Wigner resonances: $K_1(1270), K_1(1400), \Kstar(1410),\Kstar(1680)$ and $\Kstar_2(1430)$. The fit fraction of each resonance is determined and its corresponding branching fraction, given by $\BR(\Bp\to K_{res}(\to\Kp\pip\pim)\gamma)$, is computed. Furthermore, a binned maximum likelihood fit is further performed to the efficiency-corrected $m_{\kaon\pi}$ spectrum. Using the phase space decay of the three-body resonances $m_{\kaon\pi\pi}$, an efficiency map is determined and applied to the $m_{\kaon\pi}$ spectrum. The latter is modeled as the projection of two $1^-$ P-wave components, $\Kstar(892)$ and $\rho(770)$, and one $0^+$ S-wave component, $(\kaon\pi)_0^{(*0)}$. The branching fractions $\BR(\Bp\to K_{res} \pip\gamma)$ are also determined. Many of the measured branching fractions in this analysis are the first to be done or more accurate that previous world averages. Using the results of the $m_{\kaon\pi}$ spectrum, the dilution factor is  computed and yields $D_{\KS\rho\gamma}=-0.78^{+0.19}_{-0.17}$.  

\begin{figure}
\begin{center}
\includegraphics[height=6cm, width=12cm]{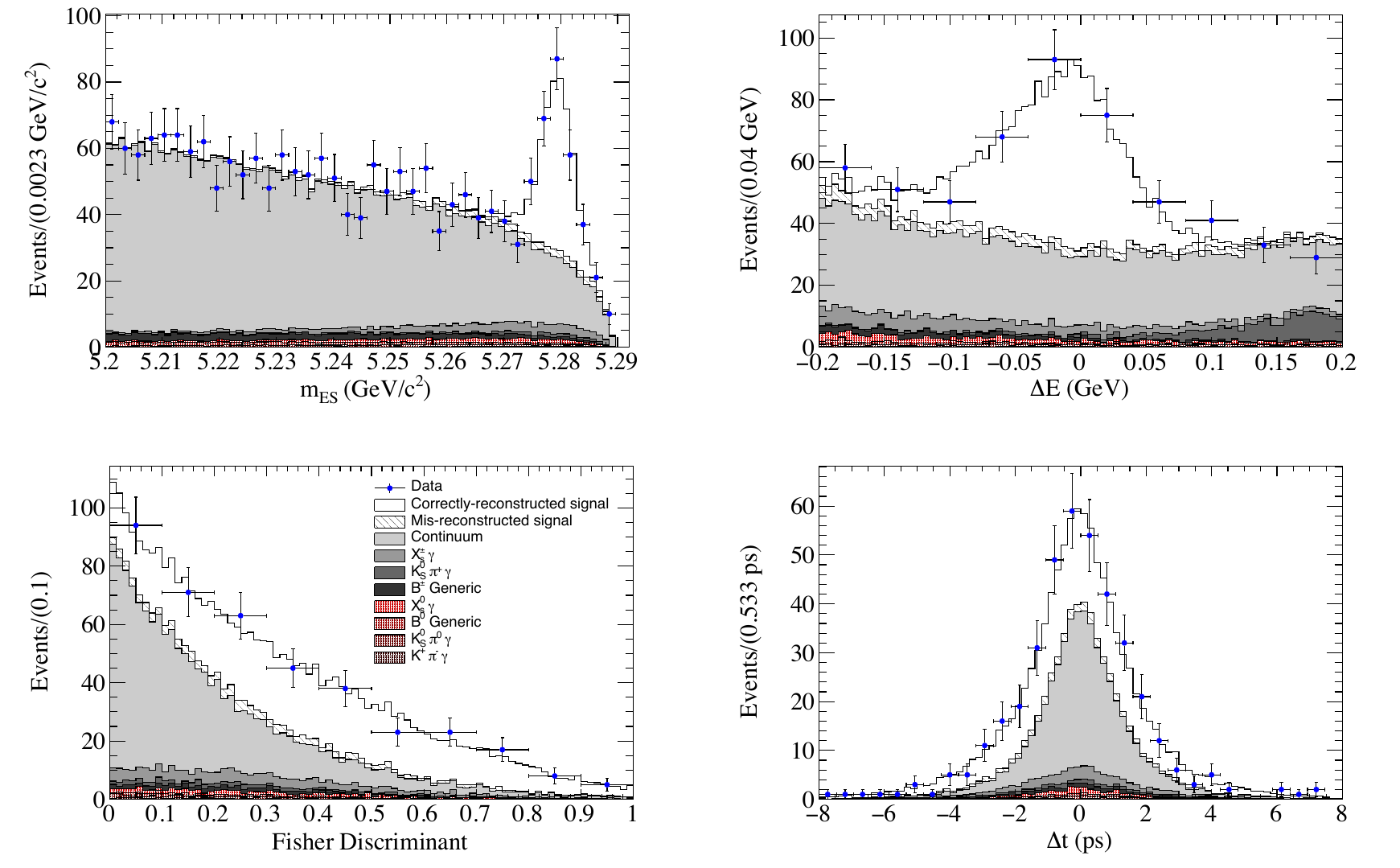}
\caption{Distributions of \mes (top left), $\Delta E$, Fisher discriminant output $F$, and $\Delta t$ with the fit results for the $\Bz\to\KS\pip\pim\gamma$ data sample. The data is shown as points with error bars and the stacked histograms represent the different background contributions.}
\label{fig10}
\end{center}
\end{figure}

 To measure the time-dependent CP asymmetry, the proper-time difference, given by $\Delta t=t_{rec}-t_{tag}$, is determined, between a fully reconstructed $\Bz\to\KS\rho\gamma$  decay ($\Bz_{rec}$) and the other \B in the event $B_{tag}$, which is partially reconstructed. The distance between the decay-vertex positions of \Btag and $B_{rec}$ is measured and transformed to $\Delta E$ using the boost, $\beta\gamma=0.56$l, of the \epem beams. To reconstruct the \Btag, a \B-flavor tagging algorithm~\cite{Btagging} is used, which combines various event variables  to achieve optimal separation between the two \B candidates in a signal event. 
$\Bz\to\KS\pip\pim\gamma$ events are reconstructed using the same signal selection as $\Bp\to\Kp\pip\pim\gamma$, but with \KS\to\pip\pim. An unbinned maximum likelihood fit is then performed to the \mes, $\Delta E$, Fisher discriminant output, $\Delta t$ and $\sigma_{\Delta t}$  distributions to extract the signal yield. The fit is shown in Fig~\ref{fig10} and  yields a branching fraction $\BR(\Bz\to\Kz\pip\pim\gamma)=(20.5\pm2.0^{+2.6}_{-2.2})\times 10^{-6}$. The CP-violation parameter is then determined to be $S_{\KS\pip\pim\gamma}=0.14\pm0.25\pm0.03$. Using the calculated dilution factor and assuming isospin asymmetry, the resulting time-dependent CP asymmetry for $\Bz\to\KS\rho\gamma$ is calculated to be: $S_{\KS\rho\gamma}=-0.18\pm0.32^{+0.06}_{-0.05}$. This measurement is in agreement with previously published results~\cite{Bellebtosg1}-\cite{Bellebtosg2} and shows no deviation from the SM prediction.

\section{Conclusion}

Various interesting and leading results are still being produced using the \babar{} dataset. An upper limit on th \Bp\to\Kp\tautau has been done for the first time. Furthermore, the angular asymmetries of \B\to\Kstar\ellell are measured and display tension with the SM expectations in the low $q^{2}$ region. In addition, the time-dependent CP-asymmetry in $\B\to\KS\rho\gamma$ has been measured and shows consistency with the SM.  \b\to\s transitions continue to be a promising probe of physics beyond the SM and a point of interest for current and future \B-factories. \\


\begin{thebibliography}{9}
\bibitem{cc}
The charge conjugate mode \Bm\to\Km\tautau is also implied.
\end{thebibliography}
\end{document}